\documentclass[prd,aps,floats,12pt]{revtex4}
\usepackage{amsmath}
\usepackage{amssymb}

\numberwithin{equation}{section}

\usepackage{epsfig}

\def\ie{{i.e.~}}
\def\beq{\begin{equation}}
\def\eeq{\end{equation}}
\def\ber{\begin{eqnarray}}
\def\eer{\end{eqnarray}}

\def\mm{{\tilde m}_P}

\def \lleq {\lower0.9ex\hbox{ $\buildrel < \over \sim$} ~}
\def \ggeq {\lower0.9ex\hbox{ $\buildrel > \over \sim$} ~}
\def\apj{{Astroph.\@ J.\ }}

\def\prl{{Phys.\@ Rev.\@ Lett.\ }}
\def\prd{{Phys.\@ Rev.\@ D\ }}

\def\plb {{Phys.\@ Lett.\@ B\ }}
\def\cqg {{Class. \@ Quant. \@ Grav.}}

\def\etal{{\it et al.}}
\def\ie {{\it ie}}
\def\n {\noindent}

\begin{document}

\title{Cosmological Hysteresis and the Cyclic Universe}

\author{Varun Sahni$^a$ and  Aleksey Toporensky$^b$}
\affiliation{$^a$ Inter-University Centre for Astronomy and Astrophysics,
Post Bag 4, Ganeshkhind, Pune 411~007, India}
\affiliation{$^b$Sternberg Astronomical Institute, Moscow State University,
Universitetsky
Prospekt, 13, Moscow 119992, Russia
}

\thispagestyle{empty}

\sloppy

\begin{abstract}
A Universe filled with a homogeneous scalar field exhibits
`{\em Cosmological hysteresis}'. Cosmological hysteresis is caused by
the asymmetry in the equation of state during expansion and contraction.
This asymmetry results in the formation of a {\em hysteresis loop}:
$\oint pdV$, whose value can be non-vanishing during each oscillatory cycle.
For flat potentials,
a negative value of $\oint pdV$ leads to the {\em increase} in amplitude
of consecutive cycles 
and to a universe with older and larger
successive cycles. 
Such a universe appears to
possess an {\em arrow of time}
even though entropy production is absent and all of the equations respect time-reversal symmetry !
Cosmological hysteresis appears to be widespread and exists
for a large class of scalar field potentials and mechanisms for
making the universe bounce.
For steep potentials, the value of $\oint pdV$ can be positive as well as negative.
The expansion factor in this case displays quasi-periodic behaviour
in which successive cycles
can be both larger as well as smaller than previous ones.
This quasi-regular pattern resembles the phenomenon of {\em beats} 
displayed by acoustic systems.
Remarkably, the expression relating
the increase/decrease in oscillatory cycles to the quantum of hysteresis 
appears to be {\em model independent}.
The cyclic scenario is extended to spatially anisotropic models and
it is shown that the anisotropy density decreases during successive
cycles if $\oint pdV$ is negative.
\end{abstract}

\maketitle


\bigskip

\section{Introduction}
\label{sec:intro}
We live in a universe that is old and very nearly 
spatially flat. The possibility that these two properties of our
universe -- its age and small spatial curvature -- could be related,
 has been the focus of considerable study in cosmology.
Tolman wondered whether a progressively older universe could be constructed
out of repeated cycles of expansion and contraction \cite{tolman}. However he was also well
aware of the fact that, for perfect fluids, the equations of motion are
reversible and so each cyclic epoch is identical to the next.
To construct a universe in which successive epochs were of longer duration
Tolman postulated the presence of a viscous fluid. 
Viscosity leads to an asymmetry in pressure during expansion and contraction
which, in turn, results in a progression of cyclic epochs of successively 
longer duration.

However Tolman did not have a prescription for avoiding the Big Bang
singularity and had to assume it a priori. An important later development which
addressed both the age issue and the flatness problem was inflation.
By driving the flatness parameter, $\Omega$, towards unity, inflation
ensured that the rapidly expanding universe, even if spatially closed,
would expand for a long duration of time. However inflation did not address
the issue of the big bang singularity 
and it has been shown that although inflation could be eternal in the
future, its past spacetime is necessarily incomplete \cite{borde}.

The present paper further develops the central idea's of an oscillatory
universe and attempts to synthesise elements of
 cyclic cosmology with the inflationary
paradigm. Extending the arguments originally proposed in \cite{nissim}
we demonstrate that a universe filled with a scalar field possesses
the intriguing property of `{\em hysteresis}'.
Cosmological hysteresis is related to the fact that the pressure
of a scalar field is usually asymmetric with respect to expansion and contraction:
$P_{\rm expansion} < P_{\rm contraction}$.
This asymmetry leads to the development of a {\em hysteresis loop},
$\oint P dV \neq 0$, during each oscillatory cycle. The loop can cause consecutive
cycles to be larger in amplitude and in duration. 
While the asymmetry between expansion and contraction is largest for
inflationary potentials, the phenomenon of cosmological hysteresis
appears to be {\em generic} and is observed also in potentials which do 
not give rise to inflation.

In \S\,2 of this paper we develop the equations which relate cosmological hysteresis
to the amplitude of successive cycles and demonstrate that these equations
 have a universal form which is independent of the scalar field potential
responsible for hysteresis. Furthermore, the presence of hysteresis appears to be robust,
and is shown to exist for quite general mechanisms of singularity
avoidance such as those predicted by Braneworld cosmology \cite{yuri} and
Loop Quantum Gravity \cite{loop,brane_loop}. 
In \S\,3 we demonstrate that $\oint P dV < 0$ for flat potentials, which
leads to the increase  in amplitude of successive cycles.
A remarkable feature of this scenario is that the universe appears to
possess an {\em `arrow of time'} even though the field equations are
formally time reversible ! 
For steep potentials the value of $\oint P dV$ can be negative as well as
positive. In this case
the phenomenon of Cosmological hysteresis
can adorn the universe with quasi-regular oscillations, or  
{\em beats}, resembling those in acoustic systems.
Section 4 discusses the behaviour of a massive scalar field during 
cosmological contraction. One finds that the 
field can grow to sufficiently large values during the contracting phase to give rise 
to a long duration inflationary phase at the commencement
of the next expansion cycle.
The behaviour of spatial anisotropy in the cyclic scenario is 
examined in \S\,5 and a brief discussion of our results is presented in \S\,6.

\section{Cosmological Hysteresis}
\label{sec:hysteresis}

An oscillatory universe requires two essential ingredients: [A] A mechanism
for singularity avoidance (when the matter density is high) and
[B] a mechanism for inducing contraction (when the matter density is low).
Below we provide a brief summary of the assumptions adopted in this paper
regarding both A and B.

{\bf [A]} {\em Cosmological Bouncing Scenario's} have been widely studied;
see \cite{rees,star79,cyclic,cyclic1,ekcyclic,graham,barrow,biswas} and \cite{novello} for a review. Within the framework of general relativity (GR)
a necessary condition for avoiding the big bang singularity is the violation
of the energy conditions usually satisfied by matter \cite{haw,cat_visser}. An alternative
viewpoint considers GR to be an `effective' theory requiring modification
when the space-time curvature becomes enormous.
The initial big bang singularity can be successfully replaced with a
`bounce' in theories incorporating both these sets of ideas, examples being
Braneworld cosmology \cite{yuri,stringy_brane}, 
Loop-quantum cosmology \cite{loop,brane_loop}, string theory motivated models 
\cite{cyclic,ekcyclic,Buchbinder,cyclic_problems,Lehners}
and other modifications to the
Einstein-Hilbert action  \cite{markov,linde,bounce1,nonlocal}. 

While a bounce in the early universe could arise in any one of the above scenario's,
the central results of this paper will be model independent and so will not depend
upon the specific mechanism sourcing the bounce.
For illustrative purposes we shall consider a bounce which is known to arise
in Braneworld cosmology (with a time-like extra dimension), in which
the Friedmann equations are modified to \cite{yuri}
\ber
H^2 &=& \frac{8\pi G}{3}\rho\left\lbrace 1 - \frac{\rho}{\rho_c}\right\rbrace
-\frac{k}{a^2}~,\nonumber\\
\frac{\ddot a}{a} &=& -\frac{4\pi G}{3}\left\lbrace (\rho+3p) -
\frac{2\rho}{\rho_c}(2\rho+3p)\right\rbrace~.
\label{eq:bounce}
\eer
This equation is also valid in Loop-quantum cosmology (LQC) when $k=0$ \cite{loop,brane_loop};
for spatially open and closed models the LQC equations are discussed in \cite{param}.
A more general class of bouncing equations which accomodates (\ref{eq:bounce}) 
and FRW dynamics as special cases is ($m \geq 1$)
\ber
H^2 &=& \frac{8\pi G}{3}\rho\left\lbrace 1 - \left(\frac{\rho}{\rho_c}\right )^m\right\rbrace
-\frac{k}{a^2}~,\nonumber\\
\frac{\ddot a}{a} &=& -\frac{4\pi G}{3}\left\lbrack (\rho+3p) -
\bigg\lbrace(3m+1)\rho + 3(m+1)p\bigg\rbrace\left(\frac{\rho}{\rho_c}\right )^m\right\rbrack~.
\label{eq:bounce0}
\eer
Eqn (\ref{eq:bounce0}) reduces to (\ref{eq:bounce}) when $m=1$, and to the FRW limit
when $\rho_c \to \infty$.
In the presence of several components contributing to the pressure and
density, one needs to replace $\rho = \sum_i\rho_i$ and $p = \sum_i p_i$
in (\ref{eq:bounce}) and (\ref{eq:bounce0}).

From (\ref{eq:bounce0}) we see that the universe bounces when
$\rho = \rho_c$, at which point $H=0$ and ${\ddot a} = 4\pi Gm(\rho_c+p_c) > 0$
(where we neglect the curvature term and assume $\rho_c +p_c > 0$). 
The prevailance of the bounce in (\ref{eq:bounce}) \& (\ref{eq:bounce0})
is not linked to the violation
of any of the energy conditions by matter, but is caused instead by a departure
of space-time dynamics from the predictions of GR
 at large values of the matter density.$^1$
\footnotetext[1]{The value of $\rho_c$ is related to fundamental
parameters appearing in Braneworld cosmology/Loop quantum cosmology.
We do not write them explicitely since the precise form of $\rho_c$
will not be required
in this paper.}
One might also note 
the following fairly general phenomenological prescription for singularity
avoidance \cite{nissim} which provides a reasonable approximation to
 (\ref{eq:bounce}) \& (\ref{eq:bounce0})
\beq
a \to a, ~\dot a
\to {}- \dot a, ~ \phi \to \phi, ~\dot \phi \to \dot \phi ~.
\label{eq:bouncetoo}
\eeq

At small values of the density ($\rho \ll \rho_c$) usually associated
with late times, higher order terms in the density
 in (\ref{eq:bounce}) \& (\ref{eq:bounce0})
can be neglected, and cosmic expansion is described by standard FRW equations
\beq
H^2 = \frac{8\pi G}{3}\sum_i\rho_i -\frac{k}{a^2}~,
~~ \frac{\ddot a}{a} = -\frac{4\pi G}{3}\sum_i(\rho_i+3p_i)~.
\label{eq:frw0}
\eeq

\bigskip
{\bf [B]} {\em Cosmological turnaround}, at late times, can take place
 in several distinct ways:

\begin{enumerate}

\item[B1.] If the Universe is spatially closed ($k=1$) and the density of
matter drops off faster than $a^{-2}$. This has been the conventional approach
of making a matter/radiation dominated universe turnaround and contract
\cite{tolman,zn83}.
Indeed for a perfect fluid such as dust ($p=0$) 
the expansion factor in a closed universe is time-symmetric and
is described by the cycloid (see figure \ref{fig:cycloid})
\begin{equation}
a(\eta) = A(1 - \cos \eta) \, , \qquad t = A(\eta -\sin \eta) \, ,
\label{eq:cycloid1}
\end{equation}
for which the space-time becomes singular at $a(t) = 0$.

\begin{figure}
\centering \psfig{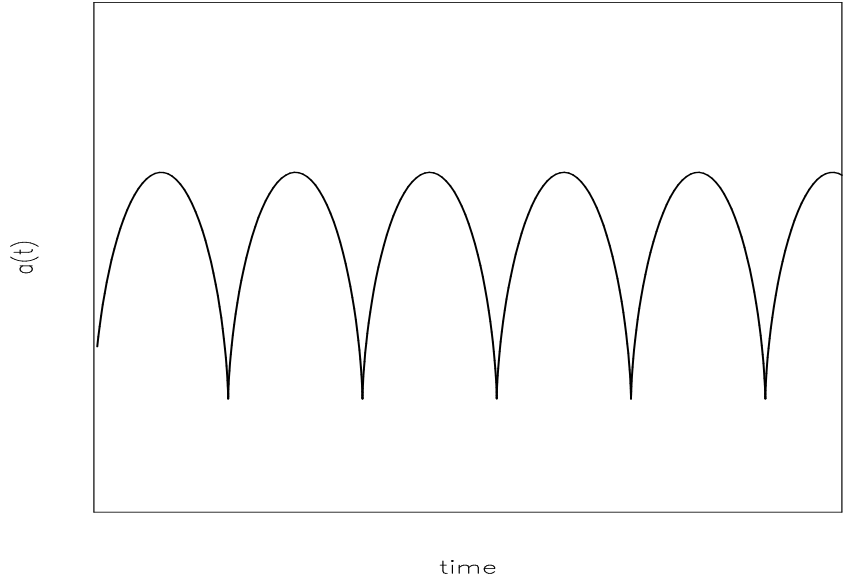}
\caption{\small The expansion factor of a spatially closed matter dominated universe
is described by the cycloid (\ref{eq:cycloid1}). }
\label{fig:cycloid}
\end{figure}

\item[B2.] The universe will turnaround if, in addition to `normal' matter 
with $\rho(t) \geq 0$, one postulates
a form of matter, $\tilde\rho$, whose energy density becomes {\em negative} at
late times. Members of this category include:

(i) $\tilde\rho(t) = - A/a^n$, with
$A>0$ and $n\leq 2$, where $n=0$  corresponds to a {\em negative}
cosmological constant $|\Lambda| \equiv A$. 
(B1 can also be regarded as being  a member of this
category for $A=k$ and $n=2$.)

(ii) Scalar fields with potentials $V_1(\phi) \propto \cos\phi$ and
$V_2(\phi) = \lambda\phi^4-m^2\phi^2$, 
allow $V(\phi)$ to evolve to negative values
as the universe expands. These models can therefore source cosmic turnaround.
Both potentials contain the possibility of giving
rise to a {\em transiently
accelerating} universe thereby providing us with interesting candidates for
dark energy \cite{alamDDE} as well as cyclic cosmology.
Other models of transient acceleration are discussed in \cite{DDE,DDE1}.

\item[B3.] A novel means of using (\ref{eq:bounce}) to obtain a cyclic
universe was suggested in \cite{freese}. These authors noted that the
density in a phantom dark energy component {\em increases} as the universe expands,
thus making the $\rho^2$ term in (\ref{eq:bounce}) relevant both at early
and at late times. In this scenario, the bounce at small values of
the expansion factor is caused by normal matter, while the universe
turns around and contracts due to phantom DE.

\end{enumerate}
The bouncing scenario [A] together with either of [B1]-[B3] 
gives rise to cyclic cosmology.

In this paper we shall assume that during some period in its history
the universe was dominated by a massive scalar field.
The presence of a scalar can make cosmological dynamics
much more versatile, as we demonstrate below. 

In this case the Lagrangian density and the energy-momentum tensor
have the form
\ber
{\cal L} &=& \frac12 g^{ij} \partial_i \phi \partial_j \phi - V(\phi) \, \nonumber\\
T_{ij}  &=& \partial_i \phi \partial_j \phi - g_{ij} {\cal L}~,
\label{eq:lagr}
\eer
and, for a homogeneous scalar field, the energy density and pressure are, respectively,
\begin{equation}
\rho = \frac12 \dot{\phi}^2 + V (\phi) \, , \qquad  p =
\frac12 \dot{\phi}^2 - V (\phi) \, , 
\label{eq:scalar_rho}
\end{equation}
and the scalar field equation of motion is
\begin{equation}
\ddot \phi + 3 H \dot \phi + \frac{dV}{d\phi} = 0 \, . 
\label{eq:scalar field}
\end{equation}
The term $3 H \dot \phi$ in (\ref{eq:scalar field}) behaves like friction
and damps the motion
of the scalar-field when the universe expands ($H>0$).
By contrast, in a contracting ($H<0$) universe, $3 H \dot \phi$ behaves like 
{\em anti-friction} and accelerates the motion of the scalar field.
Consequently a scalar field with the potential 
\cite{linde83} $V = V_0\phi^{2k}$, $k = 1,2$
displays two asymptotic
regimes \cite{belinsky}
\begin{equation} 
p \simeq - \rho \ \ \mbox{during expansion} \ \ (H > 0) \, , \label{eq:8a}
\end{equation}
\begin{equation}
p \simeq \rho \ \ \mbox{during contraction} \ \ (H < 0)  
\, . \label{eq:8}
\end{equation}

\begin{figure*}[ht]
\centerline{ \psfig{figure=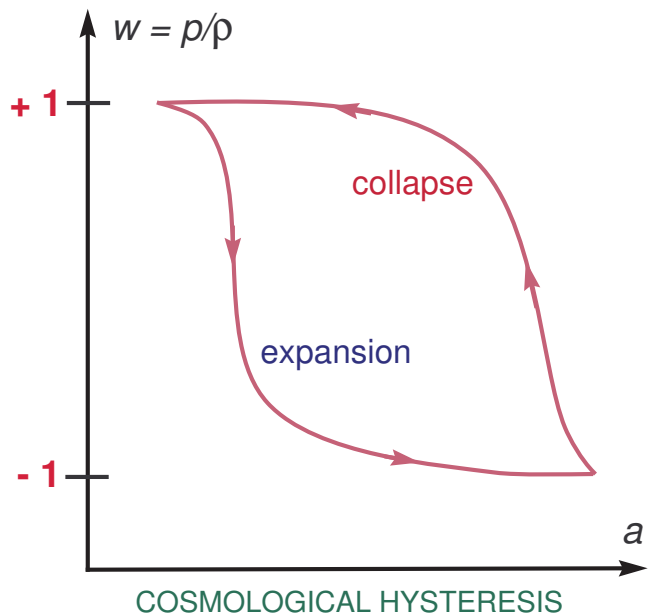,width=0.55\textwidth,angle=0} }
\bigskip
\caption{\small
An idealized illustration of cosmological hysteresis. The {\em hysteresis loop} 
shown above has $\oint p dV < 0$ evaluated over a single expansion-contraction cycle.
Rapid oscillations of
the scalar field will modify this picture, since the equation of state
 of the scalar field varies
between $-1$ and $+1$ during oscillations. Indeed, as discussed later in the text,
scalar field oscillations are indispensible for providing a non-vanishing value to the
hysteresis loop. Oscillations of the scalar during expansion scramble its phase
making all values of ${\dot \phi}$ equally likely at turnaround and leading to
$p_{\rm expansion} \neq p_{\rm contraction}$ and $\oint p dV \neq 0$.
By contrast only a unique value of ${\dot \phi}$, namely ${\dot \phi} = 0$
at {\em turnaround}, enables the scalar field to follow its original trajectory
in reverse during contraction, resulting in
$p_{\rm expansion} = p_{\rm contraction}$ and $\oint p dV = 0$. 
Consequently only a very small set of initial conditions
does not give rise to hysteresis, the latter appearing to be ubiquitous for
scalar field models which can oscillate.
}
\label{fig:hysteresis}
\end{figure*}

In the words of Zeldovich \cite{zel} ``{\em There is a moral to be learned from these
simple calculations. The result by and large conforms to Braun and Le Chatelier's
principle, which also holds in human relations:

Every system resists outside forces.

The scalar field expands, and a negative pressure (or tension) builds up.
If the expansion were created by the motion of a piston in a cyclinder
containing $\phi$, the tension would decelerate the piston. On the 
other hand, if the field is being compressed, a positive pressure builds
up, producing a force opposing the motion of the piston.''}

As Zeldovich suggests, there is nothing really surprising about the behaviour
of the scalar field in a closed universe. What is surprising,
however, is the observation \cite{nissim} that
 as the universe contracts, the scalar field does not
perform a time reversal and move back up the same
trajectory down which it descended during expansion.
What is seen instead is a lag between the trajectories describing expansion
and contraction. This behaviour, shown in figure \ref{fig:hysteresis},
is typical of systems displaying hysteresis.$^2$
\footnotetext[2]{Nature is replete with examples of hysteresis. These range from the behaviour
of ferromagnets under the influence of an external magnetic field, to
control systems, mechanics and even economics !}

Let us now discuss the effect of hysteresis on cosmological  
dynamics.
We assume, as in B2, that the
 late time behaviour of the universe is governed by the Einstein equation
\beq
H^2 \simeq \kappa \rho - \frac{A}{a^n}, 
\label{eq:late}
\eeq
where $\kappa = \frac{8\pi G}{3}$, and $A>0$, $n \leq 2$, so that
 the universe turns around
and contracts if matter satisfies the strong energy condition $\rho+3p \geq 0$.
Note that $A \equiv \Lambda < 0, ~ n=0$ corresponds to 
a {\em negative} cosmological constant,
whereas $A=1, ~n=2$ describes a universe which is spatially closed.

The scalar field (\ref{eq:lagr}) has no dissipation and therefore provides us with
an example of a perfect fluid \cite{faraoni}. However unlike other perfect fluids
such as dust or radiation,
the expansion factor for a cyclic universe filled with a scalar field need not 
display time-symmetric evolution. The reason for this is as follows.

At turnaround the universe stops expanding and begins to contract. Setting $H = 0$ in (\ref{eq:late}) we get
\beq
\kappa \rho_{t} = \frac{A}{a_{t}^n}
\label{eq:ta}
\eeq
where $\rho_{t}$ is the density and $a_{t}$ the expansion factor at turnaround.
The mass $^3$\footnotetext[3]{Our results do not depend upon
$\alpha$ in $M = \alpha \rho a^3$ and we set $\alpha = 1$ for simplicity.} 
associated with the volume $a^3$ is $M = \rho a^3$, therefore at
turnaround, 
$\kappa M_{t} = A a_{t}^{3-n}$.

\begin{figure*}[ht]
\centerline{ \psfig{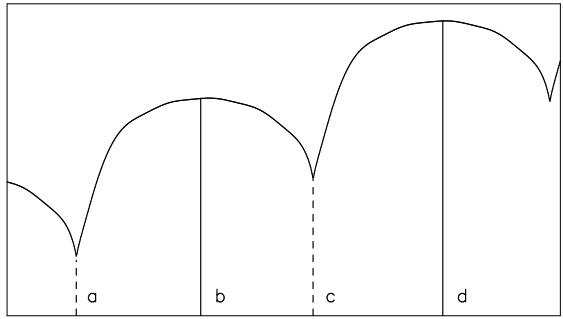} }
\bigskip
\caption{\small
The change in amplitude of successive expansion {\em maxima} is linked,
via (\ref{eq:work1}) \& (\ref{eq:work2}), to the hysteresis loop defined by
$\oint p dV := \int_b^d p dV$. By contrast, the
change in amplitude of successive expansion {\em minima} is related,
via (\ref{eq:work4}) \& (\ref{eq:work5}), to the hysteresis loop defined by
$\oint p dV := \int_a^c p dV$.
}
\label{fig:illustration}
\end{figure*}

The {\em work done} during each contraction-expansion cycle is
related to the {\em hysteresis loop}, $\oint pdV$, as follows 
\beq
\delta W = \oint pdV = \int_{\rm contraction}p~dV + \int_{\rm expansion}p~dV~.
\label{eq:work0}
\eeq
Setting $\delta W + \delta M_{t} = 0$ we get $^4$ \footnotetext[4]{The relationship
$\delta M = - p\delta V$ follows from the conservation equation
${T_i^k}_{;k}=0 \Rightarrow {\dot\rho} + 3H(\rho+p)=0$.
One might note that formulae (\ref{eq:work1}) \& (\ref{eq:work4})
agree with our numerical results for a wide range of parameters.}
\beq
-\oint pdV = \delta M_{t} = \frac{A}{\kappa} \delta a_{t}^{3-n}~,
\label{eq:work}
\eeq
from where we find the following simple expression relating the 
{\em change in amplitude} of successive cycles to the value of the
hysteresis loop ($a_{\rm max} \equiv a_{t}$)
\beq
\delta\left ( a_{\rm max}\right )^{3-n} 
\equiv \Big\lbrace a_{max}^{(i)}\Big\rbrace^{3-n} - 
\Big\lbrace a_{max}^{(i-1)}\Big\rbrace^{3-n}
= - \frac{\kappa}{A} \oint pdV~,
\label{eq:work1}
\eeq
where the hysteresis loop 
is evaluated over one complete {\em contraction-expansion}
cycle, namely
\beq
\oint pdV := \int_{a_{max}^{(i-1)}}^{a_{max}^{(i)}} pdV~,
\label{eq:work2}
\eeq
$a_{max}^{(i)}$ being the maximum value of the expansion factor
in the $i^{\rm th}$ cycle; see figure \ref{fig:illustration}.

From (\ref{eq:work1}) we find that the change in amplitude of consecutive 
cycles is sensitive both to the value of the hysteresis loop, $\oint p dV$, and the mechanism
responsible for turnaround. Two extreme cases correspond to: (i) the negative
cosmological constant ($n=0$) for which
\beq
\delta a^3_{\rm max} = - \frac{\kappa}{\Lambda} 
\oint pdV ~,
\label{eq:amax_lambda}
\eeq
(ii) the spatially closed universe ($n=2$) for which
\beq
\delta a_{\rm max} \equiv a_{max}^{(i)} - a_{max}^{(i-1)}
 = - \kappa \oint pdV ~.
\label{eq:amax}
\eeq

It is interesting that a companion relationship to (\ref{eq:work1}) can be derived
for the change in the {\em minimum value} of the expansion factor at
each successive bounce. Setting $M = \rho a^3$ as earlier, and noting
from (\ref{eq:bounce0}) that $\rho=\rho_c$ at the bounce,$^5$
\footnotetext[5]{We assume
that the role of the curvature term can be neglected at the bounce
which is a reasonable assumption provided matter satisfies the strong energy
condition, $\rho + 3p \geq 0$, near the bounce.}
we find $\delta M_{\rm bounce} = \delta(\rho_c a_{\rm min})^3$. 
The {\em work done} during an expansion-contraction cycle is 
\beq
\delta W = \oint pdV = \int_{\rm expansion}p~dV + \int_{\rm contraction}p~dV ~.
\label{eq:work3}
\eeq
Setting $\delta W + \delta M_{\rm bounce} = 0$ we get
\beq
\delta (a_{\rm min})^3 \equiv \left\lbrace a_{min}^{(i)}\right\rbrace^3 - 
\left\lbrace a_{min}^{(i-1)}\right\rbrace^3
= -\frac{1}{\rho_c}\oint pdV~,
\label{eq:work4}
\eeq
where the hysteresis loop is evaluated over one compete 
{\em expansion-contraction} cycle, namely
\beq
\oint pdV := \int_{a_{min}^{(i-1)}}^{a_{min}^{(i)}} pdV~,
\label{eq:work5}
\eeq
$a_{min}^{(i)}$ being the minimum value of the expansion factor
in the $i^{\rm th}$ cycle; see figure \ref{fig:illustration}.

Comparing (\ref{eq:work4}) and (\ref{eq:work1}) allows us to draw 
the following important
conclusions:
\begin{itemize}

\item
 From (\ref{eq:work4}) we find that the change in
the {\em minimum} value of the expansion factor depends upon the value of
the hysteresis loop, $\oint pdV$, 
and the cosmological matter density at the bounce, $\rho_c$, but is 
{\em insensitive
to the nature of turnaround}. It is important to note that an
{\em increase} in the minimum value of the expansion factor at the bounce,
determined by (\ref{eq:work4}), would lead to a corresponding {\em decrease}
in the curvature parameter $k/a_{\rm min}^2$ at the bounce,
 making it easier for a spatially
closed universe to inflate even though the curvature term may have
prevented inflation from occuring during
 earlier cycles (see also \cite{nissim,lidsey}).

\item
By contrast, 
the change in the
{\em maximum} value of the expansion factor, determined by (\ref{eq:work1}), depends upon
the nature of turnaround as well as $\oint pdV$, but
 is {\em insensitive to the density at the bounce.}
In passing one might note that the value of $n$ in (\ref{eq:late})
 is not restricted to being
an integer. Indeed if turnaround is sourced by a dynamical dark energy model
such as the scalar field with potential $V(\psi) \propto \cos{(\lambda\psi)}$ then,
as $\psi$ rolls towards the negative minimum of $V$, kinetic terms will ensure that
$n$ never quite reaches $n=0$, the value suggestive of a negative cosmological
constant. Consequently one might expect a scalar field induced turnaround to
mimic $V_0/a^{n}$ with $V_0 < 0$ and $-2\leq n < 0$. 

\item
Note that in the conventional general relativistic framework
 described by (\ref{eq:frw0})
both the 
bounce as well as turnaround can be sourced by the (positive) curvature term.
In this case
the condition for the bounce demands that, close to it, matter
{\em violates} the strong energy condition $\rho + 3p \geq 0$. On the other
hand the universe turns around if matter during much later times {\em satisfies}
the strong energy condition ! It is easy to show that in this case
(\ref{eq:work1}) is replaced by
$\delta a_{\rm max} 
 = - \kappa \oint pdV$, with $\oint pdV$ evaluated as in (\ref{eq:work2}),
while (\ref{eq:work4}) is replaced by a similar expression
$\delta a_{\rm min} 
 = - \kappa \oint pdV$, but where $\oint pdV$ is evaluated as in 
(\ref{eq:work5}). 
Within such a conventional setting bouncing models with
a scalar field were studied in \cite{aa}.  
However it seems
 unlikely that hysteresis could arise in such a scenario since $p \simeq -\rho$ 
was found to arise during contraction (inducing the bounce) and a similar equation of
state arose again during the early stages of expansion.
So the asymmetry between $p_{\rm contraction}$ and $p_{\rm expansion}$,
which is an essential requirement for hysteresis, is virtually absent in this case.

\item 
Note also that in a realistic cosmological model
only some of the matter degrees of freedom driving cosmic expansion in
(\ref{eq:bounce0}) \& (\ref{eq:frw0})
may display hysteresis. 
In this case one might expect (\ref{eq:work1}) and (\ref{eq:work4}) to
generalize to
\beq
\delta \left (a_{\rm max}\right )^{3-n} = - \frac{\kappa}{A} \oint p_\alpha dV~,~~
\delta \left (a_{\rm min}\right )^3 = -\frac{1}{\rho_c}\oint p_\alpha dV~,
\eeq
where $p_\alpha$ denotes the pressure associated with the matter component
displaying hysteresis \ie \, $p_{\alpha, {\rm expansion}} \neq p_{\alpha, {\rm contraction}}$.

\end{itemize}

\section{The\, {\em Beating}\,  Universe}

While the equations describing cosmological hysteresis appear to be
model independent, the concrete
value of the hysteresis loop $\oint pdV$ is related to the dynamics of the
scalar field and, in particular, to the form of its potential $V(\phi)$.
Let us consider some simple potentials which generate
hysteresis. 

\begin{figure*}
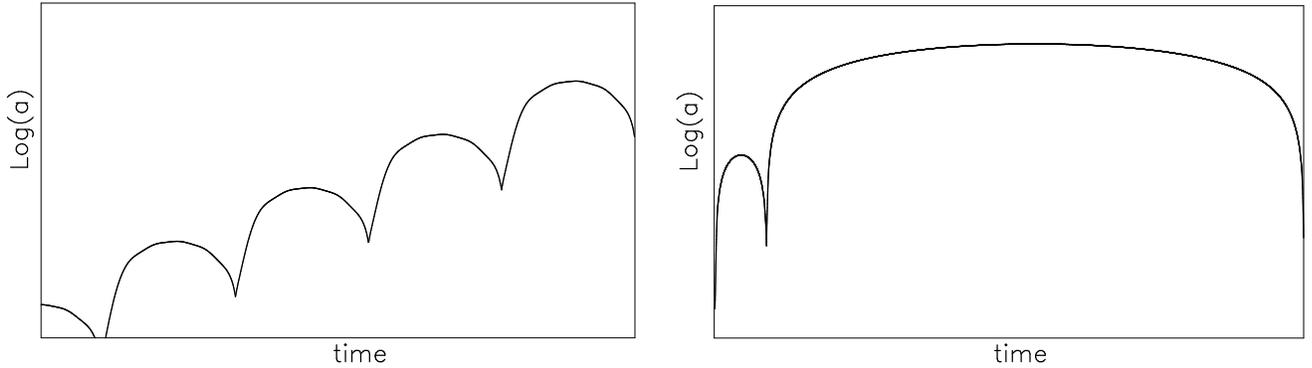

\centering
\begin{center}
\vspace{0.0cm}
$\begin{array}{@{\hspace{-0.0in}}c@{\hspace{0.2in}}c}
\multicolumn{1}{l}{\mbox{}} &
\multicolumn{1}{l}{\mbox{}} \\ [0.40in]
\epsfxsize=3.3in
\epsffile{phi2k0m05.ps} &  
\epsfxsize=3.3in
\epsffile{phi2k1m05.ps} \\
\end{array}$
\end{center}
\vspace{0.0cm}
\caption{\small The presence of hysteresis can {\em increase}
 the amplitude of successive
expansion maxima in a cyclic universe sourced by 
the 
 potential $V(\phi) = \frac{1}{2}m^2\phi^2$. In the left panel,
cosmic turnaround is caused by the presence
of a {\em negative} cosmological constant, so that $n=0$ in (\ref{eq:late}).
In the right panel, turnaround is caused by a (positive) curvature
term, so that $n=2$ in (\ref{eq:late}). In both cases the 
increased amplitude of successive cycles is described by the hysteresis
equations (\ref{eq:work1}) \& (\ref{eq:work4}).
The successively increasing expansion cycles appear to endow the
universe with an {\em arrow of time} even though the equations governing
cosmological evolution are formally time reversable and there is no
entropy production.
}
\label{fig:phi2}
\end{figure*}

{\bf [1]} The `chaotic' potential $V(\phi) = V_0\phi^{2k}$ gives rise to
inflation for $k \leq ~{\rm few}$ \cite{linde83}. 
In addition, this potential can also serve to describe `fuzzy' cold
 dark matter if $k=1$
and $V_0$ is sufficiently small \cite{fuzzy}.
It is interesting that, depending upon the value of $V_0$, this potential
can display a steady increase in the amplitude of successive cycles,
as well as other interesting features such as {\em beats} and stochasticity.
Figure \ref{fig:phi2} illustrates how the expansion factor grows with
each successive cycle when turnaround is sourced by a positive curvature
term (right panel) and a negative cosmological constant (left panel).
The issue of {\em beats} and stochasticity will be discussed slightly later
when we turn our attention to the $\cosh{(\lambda\phi)}$ potential.

A closed oscillating universe holds
 interesting consequences for the density parameter
\beq
 \Omega - 1 = (aH)^{-2} \, .
\label{eq:om}
\eeq
As we just saw, cosmological hysteresis can lead to successively
increasing expansion cycles which would
draw the value of $\Omega$ towards unity. Indeed,
 (\ref{eq:om}) clearly demonstrates that, for an identical value of $H$, larger
values of $a(t)$ will result in smaller values for $\Omega - 1$.
Clearly, a universe with {\em strong} hysteresis ($\oint pdV$ is large and negative)
 will require comparatively fewer oscillatory
cycles to reduce the value of $\Omega - 1$, as compared to one in
which hysteresis was weaker.
Thus in a universe with progressiviely increasing expansion cycles, such as the
one shown in fig. \ref{fig:phi2},
 the value of the flatness parameter will gradually be drawn
closer to unity gently ameliorating the flatness problem;
see also \cite{nissim}.

For polynomial potentials $V \propto \phi^{2k}$ the end of inflation is marked by
coherent oscillations of the scalar field and the resulting equation of state is
\beq
\langle w \rangle = \frac{k-1}{k+1}~.
\eeq
Consequently the density $\langle\rho_\phi\rangle \propto a^{-3(1+\langle w \rangle)}$ falls off
{\em faster} than dust ($\rho \propto a^{-3}$) for $k>1$, 
for instance $\langle\rho_\phi\rangle \propto a^{-4}$ in the case of the $\lambda\phi^4$
potential. As a result $M = \rho_\phi a^3$ 
is no longer a conserved quantity for $k>1$ !
Nor for that matter is $M = \rho_\phi a^3$ 
conserved during inflation (when $\rho_\phi \simeq ~constant)$
or cosmological contraction (when $\rho_\phi \propto a^{-6}$).
The formulae relating the quantum of hysteresis to the change in
expansion maxima/minima  
however remain valid since they are derived on the
basis of the conservation equations, $T{_i^k}_{;k} = 0$, which are robust
to changes in the form  of the inflationary potential and mechanisms for
making the universe turnaround and bounce.
It therefore appears that cosmological hysteresis 
extends considerably beyond the domain
of spatially closed FRW models for which it was 
originally explored  
\cite{nissim}.

{\bf [2]} Another interesting example of
cosmological hysteresis and cyclicity
 is provided by the potential (${\tilde m}_P^{-1} = \sqrt{8\pi G}$)
\beq
V = V_0(\cosh{\lambda\phi/\mm} - 1)~,
\label{eq:cosh}
\eeq
whose suitability for being a dark matter candidate was discussed in \cite{sw00,matos}.

\n
For $\lambda\phi \ll 1$, $V \propto \lambda^2\phi^2$ and the field oscillates
as pressureless matter with $\langle p\rangle = 0$. 

\n
For $\lambda\phi \gg 1$,
$V \simeq \frac{1}{2}V_0\exp{(\lambda\phi/\mm)}$ and the behaviour of the field
can be assessed using the slow roll parameters
\beq
\epsilon = \frac{\mm^2}{2}\left (\frac{V'}{V}\right )^2 \simeq \frac{\lambda^2}{2}~, ~~ 
\eta = \mm^2 \left (\frac{V''}{V}\right ) \simeq \lambda^2~.
\label{eq:slow-roll}
\eeq
Two extreme cases deserve special mention.

\begin{itemize}

\item
If $\lambda \ll 1$ then $\lbrace \epsilon, \eta \rbrace \ll 1$
and the hysteresis loop $\oint p dV$ has a large absolute value signalling 
{\em strong} hysteresis and a steady growth in amplitude of successive cycles. 

\item
For $\lambda \geq 1$, on the other hand, 
the slow-roll parameters are large and inflation with its associated
regimes (\ref{eq:8a}) \& (\ref{eq:8}) need not occur.
Surprisingly, $|\oint p dV| \neq 0$ even in this case, and the 
universe can display striking behaviour for moderate values of
the control parameter $1 \lleq \lambda \lleq ~few$.

\end{itemize}

\begin{figure*}
\centering
\begin{center}
\vspace{0.0cm}
$\begin{array}{@{\hspace{0.0in}}c@{\hspace{0.0in}}c}
\multicolumn{1}{l}{\mbox{}} &
\multicolumn{1}{l}{\mbox{}} \\ [-0.20in]
\epsfxsize=3.3in
\epsffile{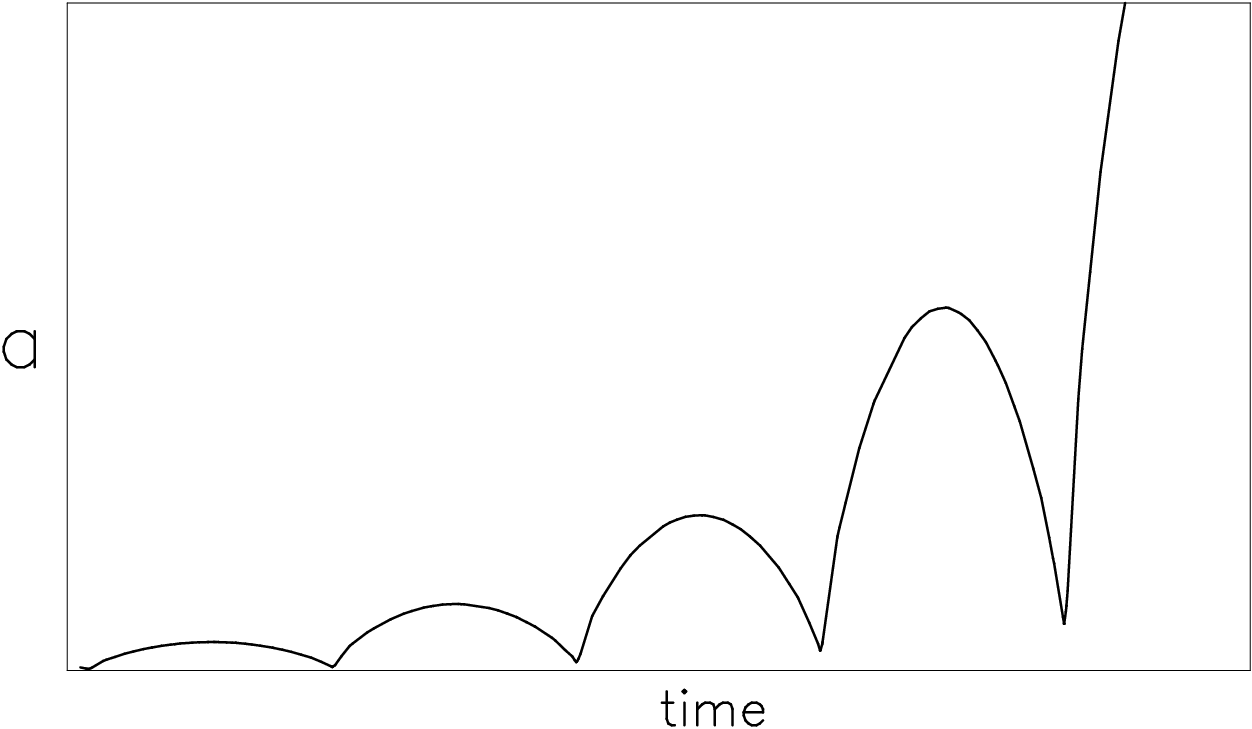} &
\epsfxsize=3.3in
\epsffile{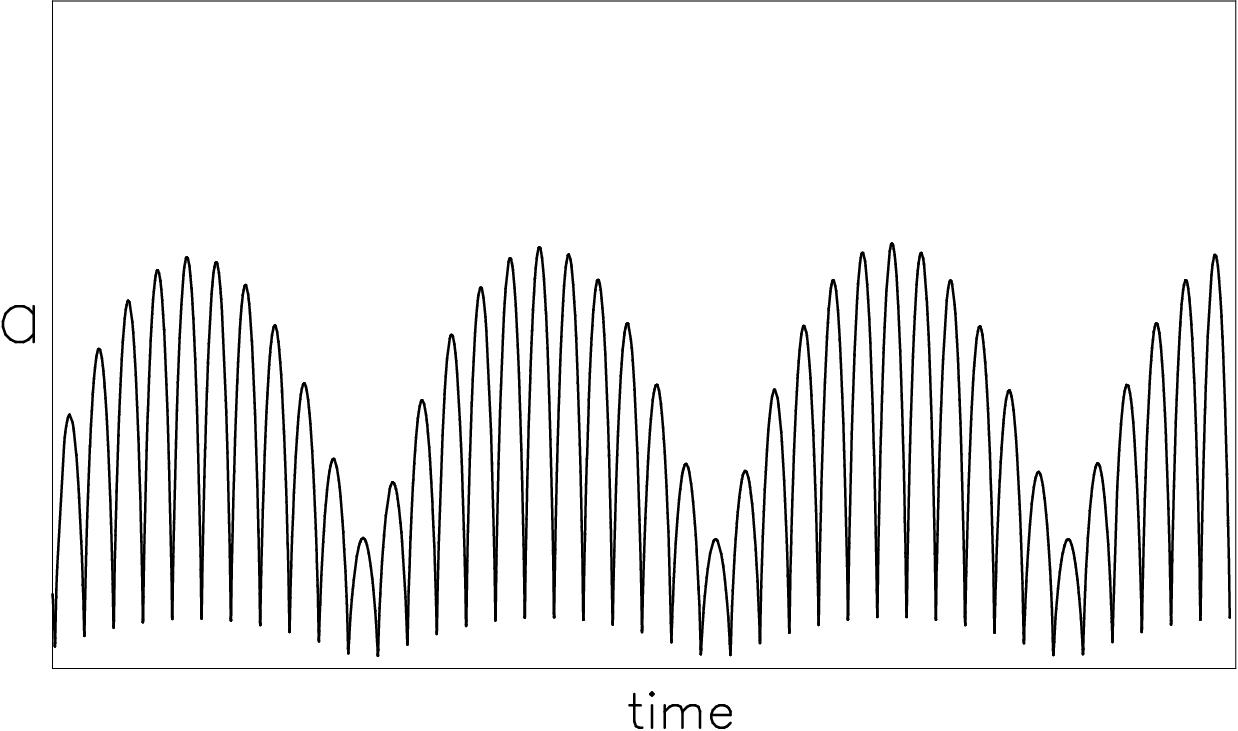} \\
\epsfxsize=3.3in
\epsffile{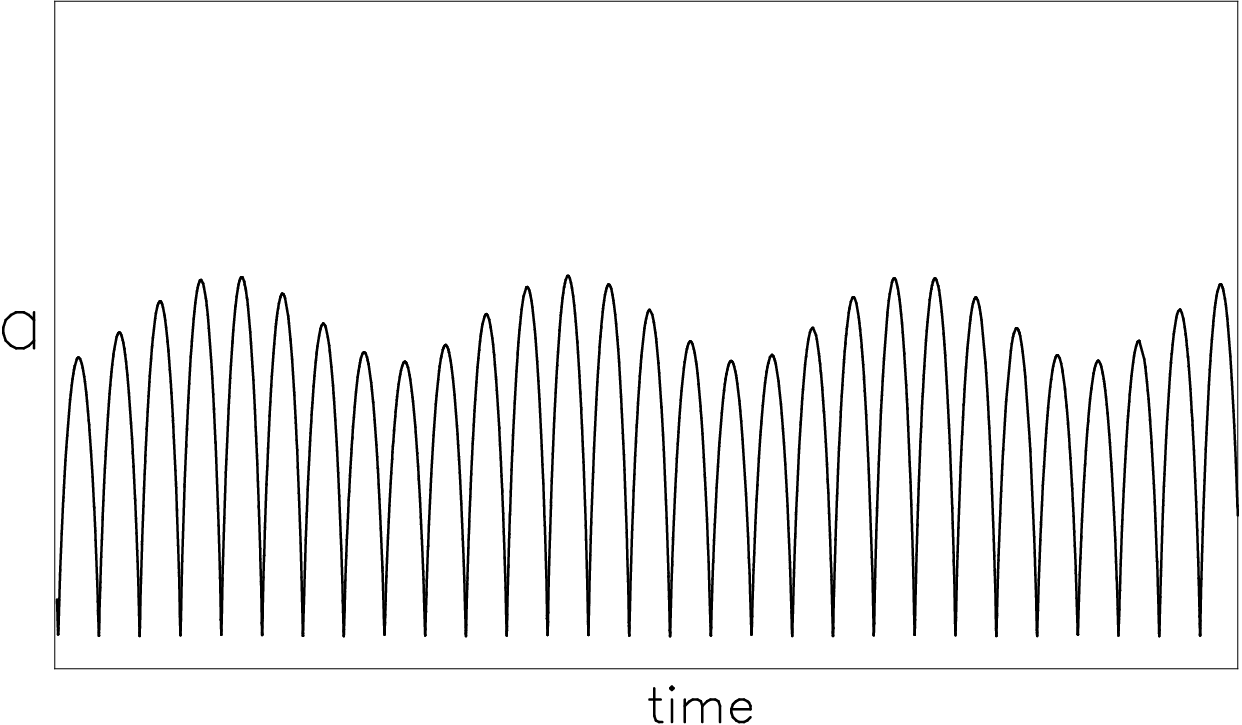} &
\epsfxsize=3.3in
\epsffile{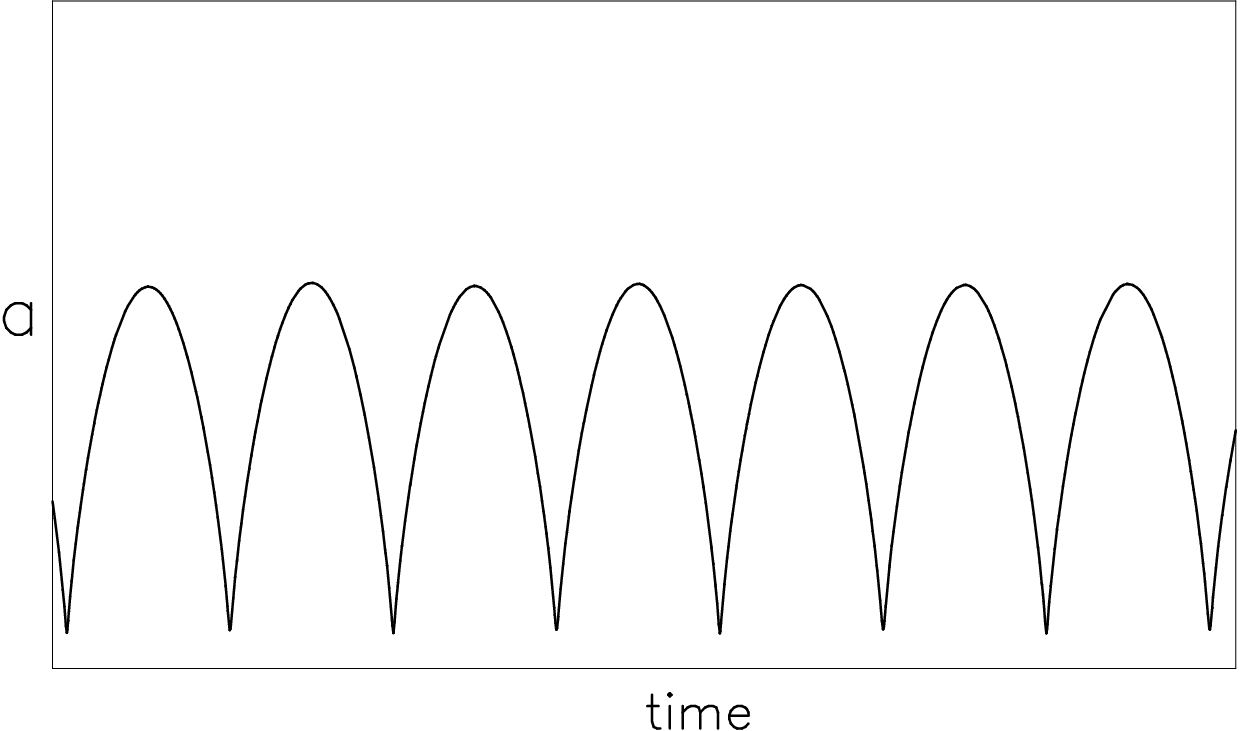} \\
\end{array}$
\end{center}
\vspace{0.0cm}
\caption{\small
A cyclic universe sourced by the
potential $V = V_0(\cosh{\lambda\phi/\mm} - 1)$. Increasing the value of the control
parameter $\lambda$ causes cyclic behaviour to change dramatically, as illustrated
in the four panels. In all cases
turnaround is caused by
a negative cosmological constant: $n=0$ in (\ref{eq:late}), while
the control (steepness) parameter has values $\lambda = 2,\, 2.5$
(upper left and right) and $\lambda = 4,\, 6$ (lower left and right).
As the control parameter increases the behaviour of the
universe undergoes a remarkable change. For $\lambda \lleq 2$
the amplitude of successive cycles increases (top left), with the increase being
larger for smaller values of $\lambda$. For moderate values of $\lambda$
(top right and bottom left panels) the universe displays a quasi-periodic
pattern reminiscent of {\em beats} in sound waves. During
{\em beats} the value of $\oint p dV$ is negative during the first
half of the larger (parent) cycle and positive during the second half.
{\rm Beats} gradually disappear as the value of $\lambda$ is
increased. The universe now begins to show 
oscillatory behaviour in which all cycles have roughly the same
amplitude and duration signifying $\oint p dV \simeq 0$ (bottom right).
In all panels the amplitude of successive cycles is governed by
the simple formulae (\ref{eq:amax_lambda}) and (\ref{eq:work4}).
}
\label{fig:cosh}
\end{figure*}

\begin{figure*}
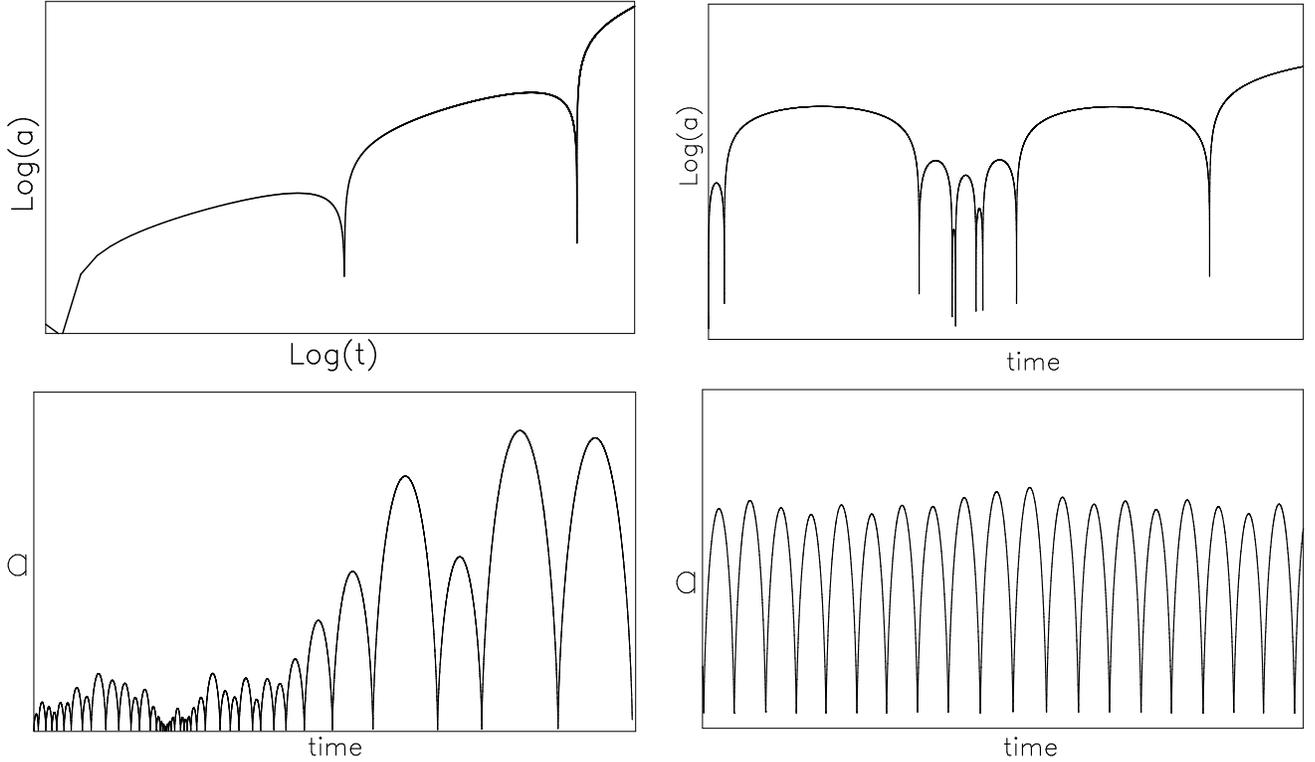

\centering
\begin{center}
\vspace{0.0cm}
$\begin{array}{@{\hspace{-0.0in}}c@{\hspace{0.2in}}c}
\multicolumn{1}{l}{\mbox{}} &
\multicolumn{1}{l}{\mbox{}} \\ [-0.20in]
\epsfxsize=3.3in
\epsffile{coshk1-5.ps} &
\epsfxsize=3.3in
\epsffile{coshk2.ps} \\
\epsfxsize=3.3in
\epsffile{coshk3-3.ps} &
\epsfxsize=3.3in
\epsffile{coshk6.ps} \\
\end{array}$
\end{center}
\vspace{0.0cm}
\caption{\small
A cyclic universe sourced by the
potential $V = V_0(\cosh{\lambda\phi/\mm} - 1)$. Increasing the value of the control
parameter $\lambda$ causes cyclic behaviour to change dramatically, as illustrated
in the four panels. In all cases
turnaround is caused by
a positive curvature term:
$n=2$ in (\ref{eq:late}), while
the control (steepness) parameter has values $\lambda = 1.5,\, 2$
(upper left and right)
and $\lambda = 3.3,\, 6$ (lower left and right).
The top two panels show the expansion factor in logarithmic units.
In the top left panel each cyclic epoch is larger in amplitude and duration
than its predecessor, indicating $\oint p dV < 0$. As the steepness parameter
is increased the behaviour of successive 
cycles becomes highly stochastic (top right and bottom left) which is indicative
of the fact that the hysteresis loop can be negative ($\oint p dV < 0$)
 as well as positive ($\oint p dV > 0$) during
cycles.
Once more we find larger amplitude cycles to have a longer duration.
Further increase of
$\lambda$ leads to a stage when hysteresis is virtually absent
and all cyclic epochs become similar.
In all panels the amplitude of successive cycles is governed by
the simple formulae (\ref{eq:amax}) and (\ref{eq:work4}).
}
\label{fig:cosh1}
\end{figure*}

The expansion factor for the $\cosh$ potential is illustrated in
figures \ref{fig:cosh} \& \ref{fig:cosh1}.
Figure \ref{fig:cosh} corresponds to the case when turnaround is sourced by
a negative cosmological constant while in fig. \ref{fig:cosh1} 
turnaround is sourced by positive (spatial) curvature.
Comparing the two figures we see similarities in the behaviour of $a(t)$
as well as interesting differences.
In both cases, moderately small values of the steepness parameter $\lambda$
give rise to a steady increase in amplitude of successive cycles.
This arises because the hysteresis loop
 $\oint p dV$ is a negative quantity 
which leads to a progressively larger universe. Note that 
while a spell of
inflation does ensure $\oint p dV < 0$,
successively larger cycles do not necessarily imply the existence of 
inflation. Indeed 
for the scalar field models which we have studied, the presence of hysteresis
appears to be a rather general phenomenon and 
all that is required for the occurance of successively larger expansion cycles is
 $\oint p dV < 0 \Rightarrow p_{\rm expansion} < p_{\rm contraction}$.
The  inflationary universe occupies the {\em extreme end} of this relationship since
$w_{\rm expansion} \simeq -1$ and  $w_{\rm contraction} \simeq 1$, as illustrated in
figure \ref{fig:hysteresis}.

As the value of $\lambda$ is increased a fundamental difference can be discerned
in the behaviour of $a(t)$ in figures \ref{fig:cosh} and \ref{fig:cosh1}.
In figure \ref{fig:cosh} (second and third panels) one sees a modulation
in the amplitude of successive cycles suggestive of {\em beats} in an acoustic
system. {\em Beats} in the expansion of the universe are characterized
by a two fold cyclic pattern with smaller duration (daughter)
 cycles nested within a large
(parent) cycle. The origin of {\em beats} can be traced to periodic changes
in the value of the hysteresis loop $\oint p dV$. During the first half of
the parent cycle, $\oint p dV < 0$, which leads to a steady
increase in the expansion maxima of successive (daughter) cycles.
As the parent cycle reaches its maximum value the hysteresis loop changes sign
so that $\oint p dV > 0$ during the next half cycle.
This leads to a steadily diminishing amplitude of daughter cycles
during the next (parent) half-cycle in accordance with (\ref{eq:work1}).
This behaviour is repeated in a self-similar manner
during subsequent parent and daughter cycles.
As in the case of acoustic beats, the daughter cycle with the smallest amplitude is
located at the boundary of two parent cycles. But unlike the acoustic case,
the modulation in the value of $a_{\rm min}$ for daughter
cycles is more graded than that in $a_{\rm max}$, with the latter showing
more pronounced changes in amplitude during a given parent cycle.

The {\em beating} universe is robust to small changes in the value of the steepness
parameter. Large values ($\lambda \ggeq {\rm few}$) however lead to diminished 
hysteresis and an end to the {\em beats} phenomenon.
In this case expansion maxima and minima equalise and the oscillatory behaviour
of the universe (fig. \ref{fig:cosh} lower right panel) begins to resemble that 
shown in figure
\ref{fig:cycloid} for a cosmological model in which expansion and contraction epochs are
identical, so that
 $p_{\rm expansion} = p_{\rm contraction}$ and $\oint p dV  = 0$.

The existence of the {\em beats} phenomenon appears to depend sensitively
on the nature of turnaround. Indeed, it appears that beats are entirely absent
if turnaround is sourced by a (positive) spatial curvature term. It is interesting
that for a closed universe the phenomenon of beats is effectively replaced by
that of {\em stochasticity}, as can readily be seen by comparing
figure \ref{fig:cosh1} with figure \ref{fig:cosh}.
During stochasticity larger expansion maxima are
interspersed with smaller ones. Such a situation arises if both
$\oint p dV < 0$ as well as $\oint p dV > 0$
can occur during consecutive cycles, and is the physical basis underlying
 {\em stochasticity}.$^6$\footnotetext[6]{The fact that stochasticity will be absent
in the cosmological model shown in figure \ref{fig:cosh} 
can be understood
from the following argument. The equations of motion for this model
(in which
recollapse is sourced by a negative cosmological constant) can be
written in terms of $\phi, {\dot\phi}, H$ 
governed by
a single constraint. Phase space is therefore two dimensional and 
stochasticity is absent due to topological reasons \cite{nonlinear}.}

The origin of stochasticity lies in the fact that for moderately steep potentials 
$\lambda \sim O(1)$,
the pressure of the scalar field during expansion
can, on occasion, exceed that during contraction. This is especially true when a 
curvature term is present, since in that case $H^2 = \kappa\rho - k/a^2$, and
the value of the Hubble parameter is smaller than it would be if $k/a^2$ were replaced
by a negative cosmological constant. This situation leads to less friction (anti-friction) during expansion
(contraction) in the scalar field equation of motion (\ref{eq:scalar field}), 
altering the nature of hysteresis and accomodating the possibility
$p_{\rm expansion} > p_{\rm contraction}$, and
$\oint pdV > 0$.
Equation (\ref{eq:amax}) then leads to $\delta a_{\rm max} < 0$, in other words
the amplitude of successive cycles can {\em decrease} as well as increase !
This is shown in figure \ref{fig:cosh1} which also informs us that
the duration of a cycle is related to its amplitude and that
larger amplitude cycles are of longer duration. 
(In the universe of figure \ref{fig:cosh}, on the other hand, all cycles are of roughly
equal duration.)
Further increase of the control parameter ($\lambda \gg 1$) causes 
the potential to steepen to such a degree that
expansion and contraction epochs become roughly similar,
so that $p_{\rm expansion} \simeq p_{\rm contraction}$. In this case hysteresis
is virtually absent, $\oint pdV \simeq 0$, 
and the amplitude of successive cycles equalizes giving $\delta a_{\rm max} \simeq 0$
(figure \ref{fig:cosh1} bottom right).

Note too that
stochasticity is usually present only if the potential is too steep 
to sustain inflation. In the presence of inflation one finds $\oint pdV < 0$,
with the result that the amplitude of successive cycles grows with time and 
stochasticity disappears.
Note too that all of the above features, namely: (i) monotonically increasing
expansion cycles (for small values of the control parameter),
(ii) {\em Beats} and stochasticity (for moderate values of the control parameter)
exist also for potentials other than $\cosh{(\lambda\phi)}$, including
$V \propto \phi^{2k}$ commonly associated with inflation.

A complementary means of generating hysteresis and the associated increase in
consecutive cycles, is by allowing the fluid filling the universe to be viscous
and dissipative \cite{tolman}. If $\zeta$ is the coefficient of  bulk viscosity then the fluid
pressure changes from its equilibrium value $p_0$ to $p = p_0 - 3\zeta H$.
Consequently $p<p_0$ during expansion while $p>p_0$ during contraction.
The resulting growth in entropy
causes sucessive expansion cycles to be larger. The possibility that bulk viscosity
might drive cosmic acceleration has been studied in \cite{bulk_visc},
and a recent attempt linking
entropy production to increased expansion cycles is discussed in \cite{biswas1}.

A central difference between this approach and ours is that our system of
equations (\ref{eq:bounce}), (\ref{eq:scalar field}) \& (\ref{eq:late})
is dissipationless and therefore formally time reversible. Yet
the presence
of cosmological hysteresis ($\oint p dV \neq 0$)
endows the universe with a plethora of new features including {\em beats},
stochasticity as well the
possibility of a regular increase in the amplitude of consecutive cycles.
The reason for this proliferation of possibility rests in the following.

As pointed out in \cite{nissim}, the presence of hysteresis is closely linked to the ability
of the field $\phi$ to oscillate.
Indeed oscillations appear to be vital for the existence of hysteresis since they
play the important role of {\em mixing} the field in phase-space
$\lbrace {\dot\phi}, \phi\rbrace$ due to which the value of $\lbrace {\dot\phi}, \phi\rbrace$
when the universe turns around and contracts is almost uncorrelated with its phase space value
when the field $\phi(t)$ began oscillating.
This phase-space mixing ensures that (during contraction) the scalar field almost never rolls up $V(\phi)$
along the same phase-space trajectory down which it descended (during expansion),
thereby ensuring 
$P_{\rm contraction} \neq P_{\rm expansion}$ and $\oint p dV \neq 0$. 

\begin{figure}
\centering \psfig{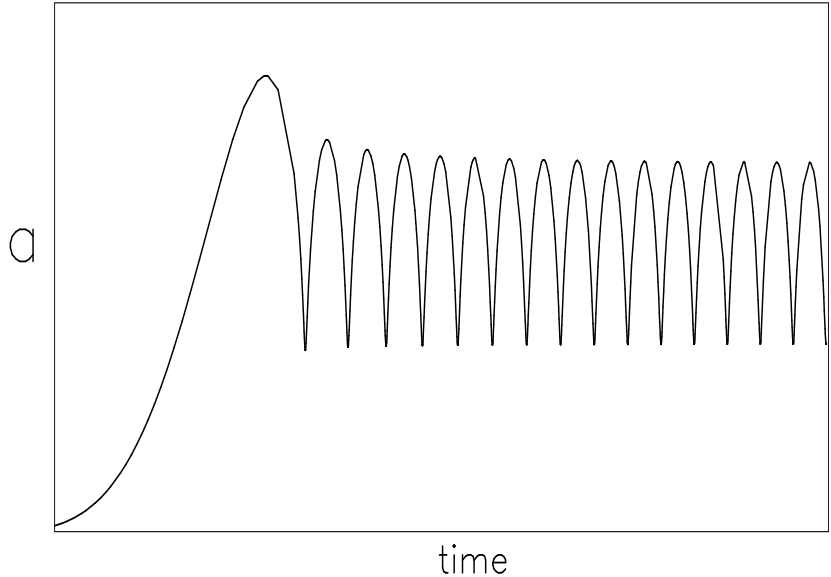}
\caption{\small The expansion factor of a spatially closed universe
dominated by a scalar field with the potential 
$V \propto \phi^{-\alpha}$, $\alpha > 0$. The
universe soon settles into an oscillatory mode in which all cycles are of equal amplitude
indicating $\oint p dV = 0$.}
\label{fig:rp}
\end{figure}

We therefore conclude that $V(\phi)$ must have a well defined minimum value
 in some region of configuration space in order to allow
the possibility of hysteresis.
Consider as an alternative the potential $V(\phi) \propto \phi^{-\alpha}$ which
does not possess a minimum and is a popular candidate for dark energy \cite{ratra}.
In this case no oscillatory phase is present which will reverse the sign
of ${\dot\phi}$ during contraction, causing $\phi(t)$ to roll up its potential instead of
down. As a result $V(\phi)$ continuously decreases in value when the universe expands
as well as contracts. 
The role of the potential during successive cycles therefore soon becomes negligible,
and cosmological dynamics becomes solely
 governed by the kinetic energy ${\dot\phi}^2 \propto a^{-6}$.
The universe, in this case, behaves as if it were dominated by a perfect fluid with
the stiff equation of state $P = \rho$ and there is no possibility of
 hysteresis, as demonstrated in figure \ref{fig:rp}.

\section{Scalar field dynamics during contraction}

An important question which arises in connection with the bouncing
scenario is whether the scalar field can rise high enough on its potential
(during contraction) to provide an adequate number of inflationary e-folds
during the ensuing round of (post-bounce) expansion.
This value can be easily estimated as follows.

As discussed earlier, the
 inflationary scalar field ($V=m^2 \phi^2/2$) in a cyclic universe
passes through three successive regimes during which:
(i) $\phi \ggeq m_P \equiv G^{-1/2}$
 and the field slow-rolls down its potential resulting in
$p \simeq -\rho$, (ii) $\phi \lleq m_P$ and the motion of the field becomes
oscillatory leading to $\langle p\rangle \simeq 0$. The oscillatory regime exists
during expansion as well as the early stages of contraction. 
(iii) During the late stages of contraction
the scalar field amplitude grows to $\phi \sim m_P$.
The field now stops oscillating and begins to climb up its potential leading to the
{\em stiff} equation of state $p \simeq \rho$. (This is in response to anti-friction
in (\ref{eq:scalar field}) since $H < 0$ during contraction.)
Below we shall provide a simple analytical estimate of how far up its potential
the field can climb before beginning its descent, soon after the universe has bounced,
and the universe starts to expand once more.

It is useful to note in this connection, the following exact expression describing
the motion of a massive scalar field
in a spatially flat universe which expands/contracts as a power law $a(t) \propto t^p$,
where $p = 2/3(1+w)$ and $w$ is the equation of state:
\beq
\frac{\phi(t)}{m_P} 
= \sqrt{\frac{t}{a^3(t)}}\bigg \lbrace AJ_\nu(mt) + BY_\nu(mt)\bigg\rbrace~,
~~ \nu = \frac{1-w}{2(1+w)}~.
\label{eq:exact}
\eeq
During the oscillatory dust-like phase, $\nu = 1/2$, and 
\beq
\phi(t) \propto \frac{\cos{(mt+\theta)}}{a(t)^{3/2}}
\label{eq:dust}
\eeq
where $\theta$ is a phase constant.
Equation (\ref{eq:dust}) informs us that the amplitude of the scalar field,
averaged over a certain number of oscillations,
increases/decreases in a contracting/expanding universe as 
$\langle \phi^2\rangle \propto a^{-3}$. 
During contraction, once the field value reaches $\phi \sim m_P$,
the oscillatory regime ceases and 
the scalar field begins to climb up its potential. 
This marks the commencement of the {\em kinetic regime} during which 
the equations of motion 
begin to be dominated by the kinetic energy of $\phi$, resulting in $w \simeq 1$, $\nu = 0$ in (\ref{eq:exact}),
and the solution
\beq
\frac{\phi(t)}{m_P}
 = AJ_0(mt) + BY_0(mt)~, ~~{\rm where}~ J_0 \simeq 1, ~Y_0 \simeq \frac{2}{\pi}\ln{(mt)}~,
~~{\rm when} ~mt \ll 1~.
\eeq
In other words
$\phi=\dot \phi_{in} t_{in}
\ln{t/t_{in}}$, where $\dot \phi_{in}$ and
$t_{in}$ are initial values of the scalar field velocity and the cosmic time
at the commencement of the kinetic regime$^7$.\footnotetext[7]{This equation can also be derived by noting that
$\ddot \phi + 3 H \dot \phi \simeq 0$ during the kinetic regime.}

Assuming that the bounce occures at the Planck energy
and remembering that $\phi \sim m_P$ when oscillations end,
we get the following result for the scalar field amplitude at the instant of the bounce
\begin{equation}
\phi_b=\frac{m_P}{\sqrt{12 \pi}} \ln{\frac{m_P}{m}}.
\label{eq:phi_bounce}
\end{equation}

(The actual value is smaller if the bounce occures at energies
below the Planck scale.)
It is worth noting that (\ref{eq:phi_bounce}) is valid for any potential which is less steep
than the exponential$^8$
\footnotetext[8]{In the derivation we used the massless approximation
which fails for steep potentials \cite{Foster}.}
and in this case the ratio
$H_B/H_{in}$ of Hubble parameters at the bounce and at the commencement
of the kinetic regime, replaces $m_P/m$ in the 
logarithm of (\ref{eq:phi_bounce}) so that
\begin{equation}
\phi_b=\frac{m_P}{\sqrt{12 \pi}} \ln{\frac{H_B}{H_{\rm in}}}.
\label{eq:phi_bounce1}
\end{equation}

While the above expression is quite general and holds for a fairly wide class of
inflationary potentials, the
number of e-folds during the post-bounce inflationary stage does depend upon the
concrete form of the inflaton potential. Consider for instance the chaotic inflationary potential,
$V=m^2 \phi^2/2$ with $m=10^{-6}m_P$, which is in excellent agreement with observations. 
Substituting $m=10^{-6}m_P$ into (\ref{eq:phi_bounce}) one obtains
$\phi_b \sim 2m_P$, and an embarassingly small value for the number of inflationary e-folds
$N=2\pi \phi_b^2 \sim 25$. Surprisingly this does not indicate
that this model can be ruled out, since, to the scalar field value at the bounce, $\phi_b$,
one needs to add $\phi_{in}$ --
the `initial' value of the scalar field at the commencement
of the kinetic regime. The value of $\phi_{in}$ depends upon the phase of the
oscillatory scalar field
 and ranges from $-m_P$ to $m_P$. 
Moreover,
immediately after the bounce the absolute value of the
scalar field continues to increase despite the fact it now
encounters a large amount of friction (the post-bounce value of $H$ being
positive). Indeed, our numerical simulations$^9$\footnotetext[9]{Although our analysis
is in broad agreement with that of \cite{cyclic} our results for ${N}$ are
somewhat smaller than those in that paper. We believe this is due to a small
typo in \cite{cyclic} due to which the value of the scalar field at the bounce
is overestimated.}
 indicate that while the value of the scalar field at the bounce is only $2.5\,m_P$,
the field manages to climb to the significantly higher value $4.6\,m_P$
soon after the bounce, 
resulting in
the observationally comfortable value $N \sim 130$.

\section{The behaviour of anisotropy in an oscillating universe}

\begin{figure}
\centering \psfig{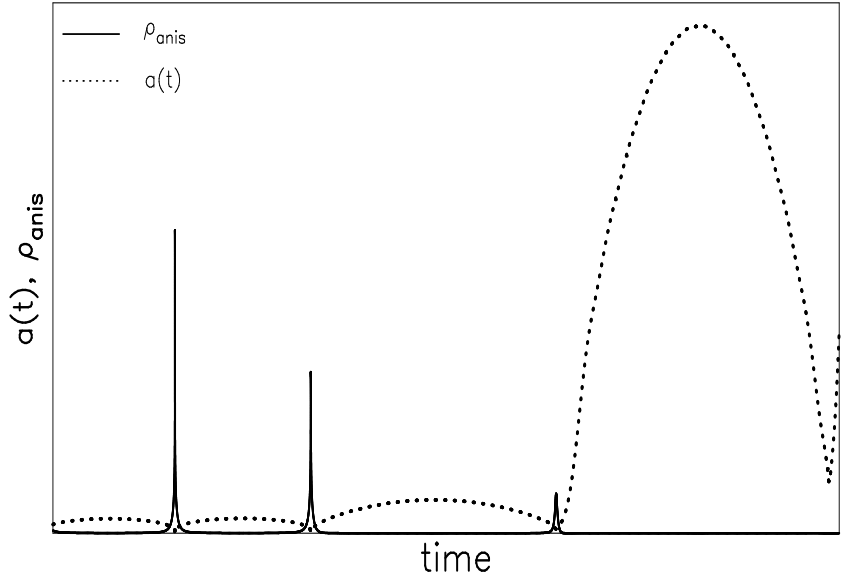}
\caption{\small The expansion factor (dotted) 
of a spatially anisotropic Bianchi I universe
dominated by a massive scalar field. The anisotropy density (solid), given by
$\rho_{\rm anis} = \Sigma/a^6$, first increases and then decreases at the time
of each bounce. The hysteresis loop for this universe is negative
$\oint p dV < 0$,  with the result that
the amplitude of successive cycles gradually increases. This leads to the
waning of 
anisotropy with each successive cycle with the result that $\rho_{\rm anis}$ 
becomes vanishingly small at the commencement of the
fifth cycle and cannot be resolved on the scale of the figure.}
\label{fig:anis}
\end{figure}

A central issue in cosmology concerns the class of initial conditions
which can give rise to a universe resembling our own.
This issue was highlighted in \cite{ch73} which showed that
isotropic models were a set of zero measure in the space of all homogenous 
solutions of the Einstein equations \cite{wald}.
Subsequently several mechanisms were identified, including cosmological particle
creation \cite{zs71}, which might successfully dissipate a large initial anisotropy within short
span of time, so that a universe which started out being highly anisotropic
would rapidly isotropize to a FRW space-time. Perhaps the most
successful of these mechanisms is inflation, which possesses a no hair
property which allows the universe to inflate from a fairly general class of
initial conditions \cite{wald1,moss}.
In the context of the present paper we would like to ask the question
as to how a large amount of anisotropy might impact the behaviour of
a cyclic universe. For this purpose 
we shall focus on a Bianchi I universe which expands at different 
rates along the three spatial directions and for which the line
element is
\beq
ds^2 = dt^2 - R_1^2(t)dx^2 - R_2^2(t)dy^2 - R_3^2(t)dz^2~.
\label{eq:bianchi}
\eeq
Introducing the directional expansion rate 
\beq
H_i = {\dot R}_i/R_i, ~~ i = 1,2,3
\eeq
and the mean expansion factor
$a(t) = (R_1R_2R_3)^{1/3}$
we find the mean expansion rate
\beq
H = \frac{\dot a}{a} = \frac{1}{3}\sum_{i=1}^3 H_i~,
\label{eq:bianchi1}
\eeq
in terms of which the expression for the anisotropy is simply
\beq
\sum_{\alpha = 1}^3 (H_\alpha - H)^2 = \frac{\Sigma}{a^6}~.
\label{eq:shear}
\eeq
The resulting (0-0) Einstein equation
\beq
3H^2 = 8\pi G\rho + \Lambda + \frac{\Sigma}{a^6}
\eeq
contains the anisotropy in the RHS as if it were an
effective energy density with the {\em stiff} equation of state
$P_{\rm anis} = \rho_{\rm anis} = \Sigma/a^6$ \cite{moss}.
In order to assess the behaviour of anisotropy in a cyclic scenario we shall assume 
$\Lambda < 0$, which permits the universe to turnaround and
contract. 
As before, we also assume that in the vicinity of the `Big Bang' 
extra-dimensional effects \cite{yuri} 
modify the FRW equations to
\ber
H^2 &=& \frac{8\pi G}{3}\rho\left\lbrace 1 - \frac{\rho}{\rho_c}\right\rbrace
+ \frac{\Lambda}{3} + \frac{\Sigma}{a^6} ~,\nonumber\\
\frac{\ddot a}{a} &=& -\frac{4\pi G}{3}\left\lbrace (\rho+3p) -
\frac{2\rho}{\rho_c}(2\rho+3p)\right\rbrace + \frac{\Lambda}{3} 
- 2\frac{\Sigma}{a^6}~.
\label{eq:bounce_anis}
\eer

Our results shown in figure \ref{fig:anis} indicate that, if the scalar field is
not too heavy ($m \lleq m_{\rm Pl}$), then the oscillating universe
displays monotonically increasing cycles which cause the anisotropy density
$\rho_{\rm anis} = \Sigma/a^6$ to decrease from cycle to cycle.
The necessary condition for the gradual disappearence of anisotropy
and the isotropisation of the universe is therefore $\oint p dV < 0$, since
this guarantees that successive cycles are larger in amplitude to previous ones.
Note that this does not require the presence of inflation, although 
a short burst of inflation at the start of every cycle
would clearly assist the  abatement of anisotropy.

\section{Discussion}
In the present paper we have shown how a cyclic universe
containing a self-interacting minimally
coupled scalar field exhibits interesting relationships,
(\ref{eq:work1}) \& (\ref{eq:work4}), linking the 
change
in the expansion factor at turnaround/bounce
 to the net work-done during a given expansion-contraction cycle: $\oint pdV$.
This relationship is quite general and requires only that the universe turnaround
and bounce, so that cyclicity is maintained. It is legitimate to ask whether such a
scenario can be realised within the framework of a universe which accelerates at
late times, and therefore resembles our own. The demand that cyclicity makes of
cosmic acceleration is that the latter be transient since otherwise
our current accelerating phase 
would become a permanent feature of our universe, preventing turnaround.
It is well known that a transiently accelerating phase can occur
in several distinct models of dark energy \cite{alamDDE,freese,DDE,DDE1}.
However, an investigation of the presence of hysteresis in such models lies
somewhat beyond the scope of the present paper and we do not discuss
it here.

Another important issue, left untouched in the present paper, concerns
the development of density inhomogeneities.
The universe today is characterized by structure which is significantly
nonlinear on scales shorter than a few Mpc.
The criterion $\delta\rho/\rho \sim 1$ marks the onset of
nonlinearity and the comoving nonlinear length scale $k_{\rm NL}^{-1}$
can be defined from the equality \cite{sc95}
\beq
\langle(\delta\rho/\rho)^2\rangle^{1/2} = D_+(t)\left (4\pi\int_0^{k_{\rm NL}}
P(k)k^2 dk\right )^{1/2} = 1
\label{eq:pert}
\eeq
where $D_+(t)$ is the growing mode of the linearized density contrast
and $P(k)$ is the power spectrum of linear density fluctuations.
For power law spectra $P(k) \propto k^n$ 
\beq
k_{\rm NL}^{-1} \propto
D_+^{2/(n+3)}(t)~,
\label{eq:kNL}
\eeq
and, since $D_+(t)$ grows with time,
it follows that the nonlinear length scale grows continuously
 from small initial
values at early times
to 
$k_{\rm NL}^{-1} \sim 10$ Mpc currently. The growth
of the nonlinear length scale with time reflects the
fact that gravitational clustering takes place
hierarchically in 
gravitational instability scenario's based on cold/warm dark matter.
The power spectrum in such scenario's is characerized by the local
spectral index 
$n_{\rm eff} = d\log{P(k)}/d\log{k}$ whose value, in the case of 
cold dark matter,
 ranges from $n_{\rm eff} \simeq 1$ on scales
$\gg  10$ Mpc to $n_{\rm eff} \simeq -3$ on scales $\ll 1$ Mpc. 
From (\ref{eq:kNL}) we find that the
growth in $k_{\rm NL}^{-1}$ will be affected both by the change in
value of the slope $n_{\rm eff}(t)$ (as successively larges scales
become nonlinear) and the behaviour of $D_+(t)$.
The latter is governed, in the absence of pressure 
and on scales significantly smaller than the Hubble length, by the Jeans-type
equation
\beq
{\ddot D_+} + 2H{\dot D_+} - 4\pi G{\bar\rho}D_+ = 0~,
\label{eq:jeans}
\eeq
from which we learn that the Hubble parameter behaves like a damping term
during expansion, when $H > 0$, causing linearized perturbations
to grow much slower than they would
 in a static universe$^{10}$\footnotetext[10]{In a realistic scenario which incorporates
both DE and turnaround,
such as \cite{alamDDE}, density perturbations would be expected to
grow as $D_+ \propto a(t)$
during matter domination, slowing considerably when the universe began to
accelerate, and speeding up once more when the universe turned around
and contracted.}: $D_+ \propto \exp{\sqrt{4\pi G{\bar\rho}}t}$. During contraction, $H < 0$, the situation
is reversed since the (negative) Hubble parameter now plays the role of an
anti-damping term which significantly speeds up gravitational
instability relative
to expansion.
This means that the nonlinear length scale will grow rapidly during contraction
eventually overtaking the comoving Hubble length $(aH)^{-1}$ which 
{\em decreases} with time in a contracting universe.
One therefore suspects that a universe resembling ours will become strongly
inhomogeneous during contraction and it is not clear whether the assumptions
which went into the derivation of the bouncing equations (\ref{eq:bounce})
 will hold in this case. (Since most bouncing solutions have been derived within the
FRW setting 
and rely on assumptions of homogeneity \cite{novello}, these concerns
are also relevant to them.)

The above argument necessarily applies to an {\em old} universe
resembling ours. It need not be true for a cyclic epoch of much shorter
duration $^{11}$. \footnotetext[11]{The above argument can also be circumvented
if, during the bounce, the universe gets `recycled' so that it embarks on its
next expansion cycle with a new set of coupling constants and a different 
inflationary potential $V(\phi)$.
A related possibility exists within the landscape paradigm if the inflaton
potential is multi-dimensional since the inflaton field $\phi$ can sample
different pieces of the potential during different cycles, as pointed out in
\cite{cyclic1}.} 
 If a shorter cyclic epoch preceeded ours then
gravitational instability would have had insufficient time to grow to
nonlinear values and the universe during contraction might persist in being
quasi-homogeneous (see \cite{perturbations} for an analysis of
linealized perturbations in cyclic models). 
Such a situation is illustrated by the right panel of
figure \ref{fig:phi2}, the first three panels of figure \ref{fig:cosh1}
 and by figure 4 of \cite{nissim}.

To summarize, we have demonstrated the presence of hysteresis 
in spatially flat and closed FRW cosmologies filled with
a self-interacting scalar field. For these models we
have established a rather simple relationship between the growth in
the expansion factor and the quantum of hysteresis, namely (\ref{eq:work1}).
We have shown that, depending upon the value of the hysteresis loop,
the universe can exhibit a wide format of behaviour including
progressively larger expansion cycles (when $\oint p dV < 0$) as well as 
stochasticity. An early inflationary epoch is not an
essential prerequisite for the existence of increasing expansion cycles.
What is required in this case is that the hysteresis loop be
negative, $\oint p dV < 0$, and even a small asymetry between the fluid
pressure during expansion and contraction can accomplish this.
In such situations, when $w_{\rm expansion} < w_{\rm contraction}$ but $w_{\rm expansion} > -1$
we sat thay hysteresis is small, in contrast to the situation portrayed in fig \ref{fig:hysteresis}
when $w_{\rm expansion} \simeq -1$ and hysteresis is large.
As demonstrated in this paper, even in situations with small hysteresis,
the moderate increase in successive cycles can draw the cosmological density
parameter towards unity, gently ameliorating the flatness problem.
The same is true for spatial anisotropy whose value declines in a universe
with growing cycles.
The presence of Cosmological hysteresis
can also adorn the universe with quasi-regular oscillations, or 
{\em beats}, reminiscent to those in acoustic systems.
One should note here that the phenomenon of beats (and stochasticity) is not
peculiar to the $\cosh{(\lambda\phi)}$ potential but has been observed in
other potentials as well, including the potential $V = \frac{1}{2}m^2\phi^2$ 
associated with chaotic inflation.

Our treatment has focussed on scalar fields possessing canonical kinetic terms
and coupling minimally to gravity. 
We find it intriguing that while the system which we study is fully
relativistic, its broad dynamical features can be enapsulated by the well
known non-relativistic thermodynamic expression $\delta E = \oint p dV$. 
It would therefore be interesting to explore 
whether the phenomenon of hysteresis is more general  and extends
to cosmologies in which some of the assumptions of this paper are relaxed, such as 
the non-canonical scalar field models associated with 
DBI inflation \cite{DBI}, k-essence \cite{kessence}, ghost condensate models \cite{ghost}, cosmologies with interacting fluid components \cite{shinji_book},  etc. 
One could also ask whether cosmological hysteresis
exists for
 scalar fields which couple non-minimally to gravity, for instance through 
$\xi R\phi^2$, Brans-Dicke and field derivative type couplings \cite{shinji}, or
in cosmological models featuring non-local gravity \cite{nonlocal}.
We also leave untouched the interesting issue of quantum stability of a cyclic
model displaying hysteresis, which touches on issues beyond the scope
of the present paper \cite{vilenkin}.
Finally, it may also be worth enlarging the present analysis to 
(higher-dimensional) anisotropic models in which some of the spatial directions
expand while others contract, and ask whether a form of hysteresis
might exist in this case too.

\section*{Acknowledgments}
We acknowledge useful discussions with Tarun Saini, Yuri Shtanov,
Parampreet Singh and Sanil Unnikrishnan.

\end{document}